\documentclass[twocolumn,prl,showpacs,preprintnumbers,amsmath,amssymb]{revtex4-1}
\usepackage{geometry}
\usepackage{color}
\usepackage{graphicx}
\begin{document}


\title{Angle-resolved photoemission spectroscopy of the low-energy electronic structure of superconducting Pr$_2$CuO$_4$ driven by oxygen non-stoichiometry}

\author{M. Horio$^1$, Y. Krockenberger$^2$, K. Koshiishi$^1$, S. Nakata$^1$, K. Hagiwara$^1$, M. Kobayashi$^3$, K. Horiba$^3$, H. Kumigashira$^3$, H. Irie$^2$, H. Yamamoto$^2$, and A. Fujimori$^1$}

\affiliation{$^1$Department of Physics, University of Tokyo, Bunkyo-ku, Tokyo 113-0033, Japan}
\affiliation{$^2$NTT Basic Research Laboratories, NTT Corporation, Atsugi, Kanagawa 243-0198, Japan}
\affiliation{$^3$KEK, Photon Factory, Tsukuba 305-0801, Japan}

\date{\today}

\begin{abstract}
Bulk crystals of electron-doped cuprates with the T'-type structure require both Ce substitutions and reduction annealing for the emergence of superconductivity while the reduction annealing alone can induce superconductivity in thin films of the T'-type cuprates. In order to reveal low-energy electronic states which are responsible for the superconductivity, we have conducted angle-resolved photoemission spectroscopy measurements on thin films of the superconducting Ce-free T'-type cuprate Pr$_2$CuO$_4$. The results indicate that the overall band structure and the Fermi surface area of the superconducting Pr$_2$CuO$_4$ are similar to those of superconducting Ce-doped bulk single crystals, highlighting the importance of the actual electron concentration rather than the Ce concentration when discussing the physical properties of the T'-type cuprates.
\end{abstract}


\maketitle

High-temperature superconductivity in the cuprates emerges when carriers are doped into an antiferromagnetic (AFM) Mott-insulating parent compound through chemical substitutions. In the case of the electron-doped cuprates $Ln_{2-x}$Ce$_x$CuO$_4$ ($Ln$: rare earth) with the so-called T'-type structure, where Cu atoms are coordinated by four O atoms in the square-planar configuration, electrons are doped through the substitution of Ce$^{4+}$ for $Ln^{3+}$. Recently, the electronic structure of the T'-type cuprates has become a subject of intense debate because superconductivity was realized without Ce substitution in thin-film samples \cite{Tsukada2005a,Matsumoto2009b,Krockenberger2013}, which challenges the common belief that the parent cuprates are Mott insulators. An indispensable treatment to realize the superconductivity is post-growth reduction annealing, whose role is believed to remove a small amount of impurity oxygen atoms at the apical sites \cite{Radaelli1994,Schultz1996} which act as strong scattering centers \cite{Xu1996}. Owing to the larger surface-to-volume ratio of the thin films than bulk, apical oxygen atoms are expected to diffuse more efficiently towards the surface than in bulk crystals, which may lead to the emergence of superconductivity in thin films even without Ce substitution \cite{Naito2016}. On the theoretical side, studies using the local density approximation plus dynamical mean-field theory have classified the parent compounds of the T'-type cuprates as weakly correlated Slater insulators or metals rather than Mott insulators \cite{Das2009,Weber2010a,Weber2010b}.

Primary effects of annealing on the physical properties of the T'-type cuprates are to suppress the scattering rate of quasi-particles \cite{Xu1996,Higgins2006} and to reduce both the AFM correlation length and the N\'{e}el temperature of the AFM order competing with superconductivity \cite{Mang2004a,Adachi2016}. This is probably rendered by the removal of apical oxygen atoms. In addition, it has been revealed by recent angle-resolved photoemission spectroscopy (ARPES) studies on Ce-doped bulk single crystals \cite{Horio2016,Song2017} and Ce-free insulating thin films \cite{Wei2016} that reduction annealing also leads to electron-doping. Here, the electron concentrations have been estimated from the Fermi surface areas. The increase of the electron concentration with reduction annealing could be explained if the oxygen atoms are removed not only from the apical sites but also from the regular sites (the O sites in the CuO$_2$ planes and/or the $Ln_2$O$_2$ layers) in the course of annealing. The possibility of oxygen-vacancy creation at the regular sites by reduction annealing has indeed been pointed out by infrared and Raman spectroscopy studies on bulk crystals and thin films \cite{Riou2004,Richard2004}, as well as by a transport study \cite{Yu2007}. Provided that the electron concentration is not solely determined by the Ce concentration but also by oxygen vacancies, the band filling of the superconducting (SC) Ce-free T'-type cuprates should be directly determined. Horio \textit{et al.} \cite{Horio2017a} have recently reported hard x-ray photoemission and soft x-ray absorption spectroscopy measurements on the SC Ce-free T'-type Nd$_2$CuO$_4$ thin films. By reduction annealing and thus inducing superconductivity in Nd$_2$CuO$_4$ thin films, chemical potential was shifted to the same level as that of 15-19 \% Ce-doped SC films, suggesting a considerable amount of electron doping by annealing. While these core-level spectroscopies probe the local, element-specific electronic structure precisely, they do not provide momentum-resolved low-energy information unlike ARPES such as (i) Fermi surface area which enables the quantitative evaluation of electron concentration \cite{Horio2016,Song2017,Wei2016}, (ii) AFM correlations appearing as a pseudogap at ``hot spots'' \cite{Richard2007,Song2012,Horio2016,Song2017}, and (iii) hopping parameters which characterize the band structure and significantly affect the superconductivity in cuprates \cite{Ikeda2009,Sakakibara2010}. All the information is desired for the establishment of the phase diagram. However, ARPES measurements, which are suitable for that purpose, are highly surface sensitive and have not been applicable to reduction-annealed thin films because film surfaces are contaminated once the film is taken out of a growth chamber for post-growth {\it ex-situ} annealing \cite{Krockenberger2013}.

Recently, Krockenberger {\it et al}.~\cite{Krockenberger2015} successfully grew SC thin films of Ce-free T'-type cuprate Pr$_2$CuO$_4$ without post-growth annealing. Those films were grown on a perfectly lattice-matched substrate under relatively stronger reducing condition on the verge of phase decomposition. Note that the ``relatively stronger reducing condition" means higher temperature and lower ozone pressure than in a typical growth, and does not mean the usage of reducing agents such as CaH$_2$. The grown films were capped with a Se protective layer to prevent surface contamination during the transfer to the ARPES apparatus in air. The Se layer was successfully removed {\it in vacuo} prior to the ARPES measurements, as we shall describe below in detail. In this Letter, we report the ARPES studies on thus prepared thin films of SC Pr$_2$CuO$_4$ to reveal the low-energy electronic structure relevant to the superconductivity. We have measured both SC and non-SC Pr$_2$CuO$_4$ thin films grown under different reducing conditions, and determined their carrier concentrations quantitatively from the area of the observed Fermi surface. The results indicate that the band filling of SC Pr$_2$CuO$_4$ is larger than half-filling and is comparable to that of SC Ce-doped bulk single crystals.

\begin{figure}
\begin{center}
\includegraphics[width=60mm]{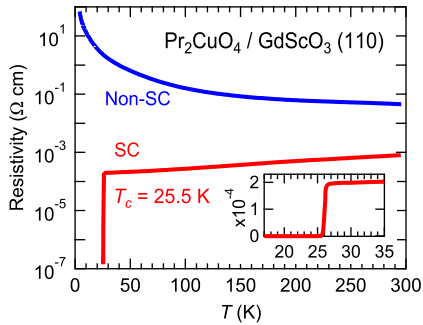}
\end{center}
\caption{Resistivity of the superconducting (SC) and non-SC Pr$_2$CuO$_4$ thin films studied in the present work plotted against temperatures. The inset shows the temperature range near the SC transition.}
\label{PCO_rhoT}
\end{figure}

Pr$_2$CuO$_4$ thin films for the ARPES measurements were synthesized at NTT Basic Research Laboratories using the molecular beam epitaxy method on GdScO$_3$ (110) substrates. By varying the growth condition as described elsewhere \cite{Krockenberger2015}, two Pr$_2$CuO$_4$ films were grown: one was non-SC with the $c$-axis length of 12.207 \AA\, and the other was SC with the $c$-axis length of 12.190 \AA. The SC film exhibited $T_\mathrm{c}$ of 25.5 K after the ARPES measurement as shown in Fig.~\ref{PCO_rhoT}. The SC film grown under a stronger reducing condition has a shorter $c$-axis length than the non-SC film, suggesting that less apical oxygen atoms are incorporated. ARPES measurements were carried out at beamline 2A of Photon Factory using circularly polarized $h\nu = 55$ eV photons. The total energy resolution was set at 35 meV. Just prior to the ARPES measurements, x-ray photoemission measurements were also performed on the same sample surfaces without moving the sample position to check the surface condition with linearly polarized $h\nu = 1200$ eV photons. All the measurements were performed at 15 K under a vacuum better than $7 \times 10^{-11}$ Torr.

\begin{figure}
\begin{center}
\includegraphics[width=75mm]{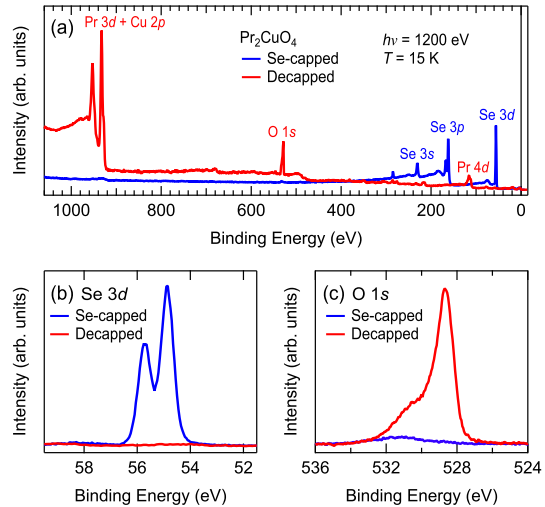}
\end{center}
\caption{X-ray photoemission spectra of the SC Pr$_2$CuO$_4$ thin film before and after removing the Se-capping layer. (a) Survey scan over a wide energy range. (b),(c) Se 3$d$ and O 1$s$ spectra. An integral-type background has been subtracted from each spectrum.}
\label{Se_cap}
\end{figure}

In order to keep the surfaces of the Pr$_2$CuO$_4$ thin films clean until the ARPES measurements, Se capping was employed. After the growth, the films were rapidly cooled down to 70 $^\circ$C and amorphous Se was evaporated onto the film surfaces up to the thickness of $\sim 50$ nm under the vacuum of $\sim 1 \times 10^{-9}$ Torr to protect the film surface from contamination before taken out of the growth chamber. The films were transferred in air from NTT Basic Research Laboratories to Photon Factory. Then, the films were heated inside the preparation chamber at 150 $^\circ$C for 30 minutes under a vacuum better than $2 \times 10^{-9}$ Torr to desorb the Se cap, and transferred {\it in vacuo} to the measurement chamber. Although the Se-capping method is now widely used for vacuum-ultra-violet (VUV) APRES measurements on thin films which contains Se such as the topological insulator Bi$_2$Se$_3$ \cite{Xu2012} and the iron-based superconductor FeSe \cite{Liu2012}, it is not trivial if it also works for oxides such as Pr$_2$CuO$_4$. To examine the surface condition, we first measured core-level photoemission spectra using soft x rays. Figure~\ref{Se_cap}(a) shows x-ray photoemission spectra in a wide energy range before and after the decapping. While the spectrum was dominated by Se core-level peaks before decapping, once the sample was heated, core-level peaks of Pr, Cu, and O emerged. The intense Se 3$d$ peaks became indiscernible after heating [Fig.~\ref{Se_cap}(b)], indicating that the desorption of amorphous Se was almost complete. The emerging O 1$s$ core level had a sharp peak and a long tail on the higher binding energy side. The intensity of the shoulder at $\sim$ 531 eV is comparable to that of Nd$_{2-x}$Ce$_x$CuO$_4$ bulk single crystals cleaved {\it in-situ} in a previous soft x-ray photoemission study \cite{Suzuki1990a}. Therefore, the Se cap seemingly protected the surface of Pr$_2$CuO$_4$ from contamination rather efficiently. It is possible that the Se-capping method also works for other oxides and extends the applicability of ARPES to oxide thin films.

\begin{figure}
\begin{center}
\includegraphics[width=85mm]{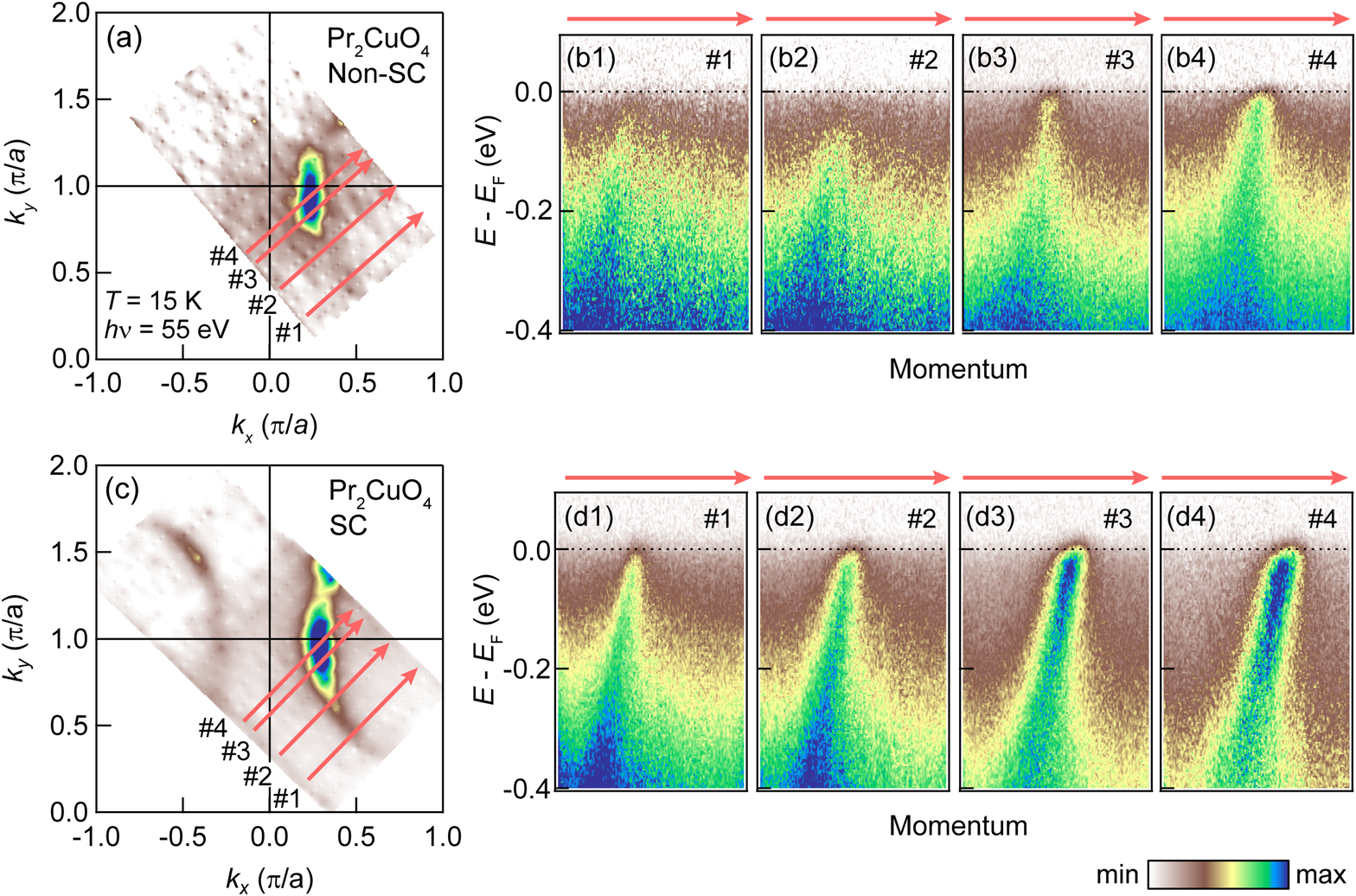}
\end{center}
\caption{ARPES spectra of Pr$_2$CuO$_4$ thin films. (a) Fermi surface of non-SC Pr$_2$CuO$_4$. (b1)-(b4) Band images taken along the cuts indicated in (a). The intensity becomes weaker as the band approaches the Fermi level ($E_\mathrm{F}$). In particular, the band does not reach $E_\mathrm{F}$ for cut \#2, indicating the opening of the AFM pseudogap at the hot spot. (c) Fermi surface of SC Pr$_2$CuO$_4$. (d1)-(d4) Band images taken along the cuts indicated in (c). The intensity remains strong up to $E_\mathrm{F}$ for every cut, indicating the strong suppression of the AFM pseudogap.}
\label{PCO_FS}
\end{figure}

Immediately after confirming the surface condition, we started ARPES measurements using VUV photons. Fermi surfaces and band dispersion were successfully observed for both SC and non-SC Pr$_2$CuO$_4$ thin films as shown in Fig.~\ref{PCO_FS}. For the non-SC film, in going from the node [cut~\#1:~Fig.~\ref{PCO_FS}(b1)] to the hot spot [cut~\#2:~Fig.~\ref{PCO_FS}(b2)], where the Fermi surface and the AFM Brillouin-zone boundary cross each other, the band is gapped. This gap is characteristic of electron-doped cuprates called an AFM pseudogap, which originates from band folding due to AFM short-range order \cite{Armitage2001b,Matsui2005a,Park2007,Park2013}. Approaching the antinode, the upper part of the split band is lowered, crosses $E_\mathrm{F}$, and produces significant spectral intensity just below $E_\mathrm{F}$ [cut~\#3:~Fig.~\ref{PCO_FS}(b3)]. The resulting Fermi surface shown in Fig.~\ref{PCO_FS}(a) is discontinuous: clearest in the antinodal segment and barely discernible in the nodal segment. For the SC sample, on the other hand, spectral intensity at the hot spot is not lost [Fig.~\ref{PCO_FS}(d2)] and the entire Fermi surface remains connected [Fig.~\ref{PCO_FS}(c)], suggesting that the AFM pseudogap is strongly suppressed (See Supplemental Material \cite{Suppl}, which includes Refs.~\cite{Kaminski2004,Matt2015}, for detailed analyses of the AFM pseudogap using energy distribution curves). The weakening of the AFM pseudogap by reduction annealing has been reported in several ARPES studies on Ce-doped compounds \cite{Richard2007,Song2012,Horio2016,Song2017}, but it is remarkable that robust antiferromagnetism in the parent T'-type cuprates is indeed suppressed without Ce substitution and only by oxygen reduction.

\begin{figure}
\begin{center}
\includegraphics[width=85mm]{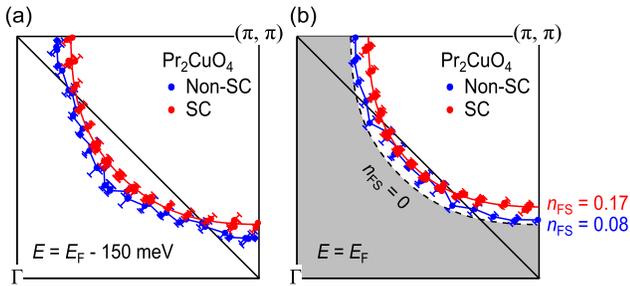}
\end{center}
\caption{Constant-energy surfaces and Fermi surfaces of Pr$_2$CuO$_4$ thin films. (a) Peak positions of MDCs at $E = E_\mathrm{F} - 150$ meV integrated within the energy window of $\pm 10$ meV for the SC (red dots) and non-SC Pr$_2$CuO$_4$ (blue dots) films. The red curve represents the constant energy surface at $E = E_\mathrm{F} - 150$ meV fitted to the tight-binding model. (b) $k_\mathrm{F}$ points of the SC (red dots) and non-SC (blue dots) films determined as described in the text. $k_\mathrm{F}$ points of the SC films are fitted to the tight-binding model (red solid curve). The area of the fitted Fermi surface yields the electron concentration $n_\mathrm{FS}$ of 0.17 and 0.08 for the SC and non-SC films, respectively. For comparison, the Fermi surface for $n_\mathrm{FS} = 0$ (dashed black curve) for the same tight-binding model is also plotted. Gray shaded region is occupied when $n_\mathrm{FS} = 0$.}
\label{PCO_nFS}
\end{figure}

In order to estimate the electron concentration of the Pr$_2$CuO$_4$ thin films, the constant-energy surfaces at 150 meV below $E_\mathrm{F}$ were determined from the peak positions of the momentum distribution curves (MDCs) at $E = E_\mathrm{F}-150$ meV (See Supplemental Material \cite{Suppl} for the original images of the constant-energy surfaces) and are plotted in Fig.~\ref{PCO_nFS}(a). While the Fermi surface was truncated by the AFM order for the non-SC film [see Fig.~\ref{PCO_FS}(a)], the constant-energy surface at the high binding energy plotted in Fig.~\ref{PCO_nFS}(a) remains connected and is less affected by the AFM order. Richard {\it et al.}~\cite{Richard2007}, however, demonstrated that the constant-energy surface of Pr$_{2-x}$Ce$_x$CuO$_4$ bulk single crystals at $E = E_\mathrm{F}-100$ meV is identical between as-grown and annealed samples, and concluded that the electron concentration of bulk single crystals is not affected by moderate annealing. In the case of the present Pr$_2$CuO$_4$ thin films the hole-like surface centered at ($\pi$, $\pi$) is obviously smaller for the SC film, indicating that the electron concentration is larger for the SC film synthesized under the more strongly reducing condition.

In Fig.~\ref{PCO_nFS}(b), Fermi surfaces are plotted for both SC and non-SC Pr$_2$CuO$_4$ thin films. For the SC sample, $k_\mathrm{F}$ positions were determined from the peaks in the MDCs at $E_\mathrm{F}$. As for the non-SC sample, the same method was used except for around the node and hot spot, where MDC-peak positions from $E-E_\mathrm{F} =$ -70 meV to -20 meV were extrapolated to $E_\mathrm{F}$ (See Supplemental Material \cite{Suppl} for more details about the $k_\mathrm{F}$ determination). The obtained Fermi surface of the non-SC thin film is not smooth around the hot spot, reflecting the presence of the AFM pseudogap. The area of the Fermi surface of the non-SC Pr$_2$CuO$_4$ corresponds to the actual electron-doping level $n_\mathrm{FS}$ of 0.08 $\pm$ 0.03, which is larger than the half-filling ($n_\mathrm{FS} = 0$). This $n_\mathrm{FS}$ value is similar to that of an insulating T'-type La$_2$CuO$_4$ thin film ($0.09 \pm 0.02$) obtained in a recent ARPES study \cite{Wei2016}.

For the SC Pr$_2$CuO$_4$ thin film, as displayed in Fig.~\ref{PCO_nFS}(b), the Fermi surface is satisfactorily fitted to the tight-binding model,
\begin{align}
\epsilon - \mu =& \epsilon _0 - 2t (\mathrm{cos} \ k_xa + \mathrm{cos} \ k_ya)  \nonumber \\ 
 & -4t'\mathrm{cos} \ k_xa \ \mathrm{cos} \ k_ya -2t"(\mathrm{cos} \ 2k_xa + \mathrm{cos} \ 2k_ya),
\end{align}
where $t$, $t'$, and $t''$ denote the transfer integrals between the nearest-neighbor, second-nearest-neighbor, and third-nearest-neighbor Cu sites, respectively, and $\epsilon_0$ denotes the energy of the band center relative to the chemical potential $\mu$. Fixing $t''/t'$ at -0.50 in the course of the fitting, we obtained best fit for $t'/t = -0.22$ and $\epsilon_0/t = -0.10$. From the fitted Fermi surface, the electron-doping level $n_\mathrm{FS}$ was estimated to be 0.17 $\pm$ 0.02, which is even larger than that of the non-SC Pr$_2$CuO$_4$ thin film, consistent with the tendency found in the constant-energy surfaces at $E = E_\mathrm{F} - 150$ meV. The experimentally determined hole-like Fermi surfaces are obviously smaller than the Fermi surface at half-filling [$n_\mathrm{FS} = 0$ as shown in Fig.~\ref{PCO_nFS}(b)], and hence one can conclude that the SC Pr$_2$CuO$_4$ thin film has a larger amount of electrons in the CuO$_2$ planes than at half-filling. The $n_\mathrm{FS}$ value obtained for the SC film is comparable to the Ce concentration $x$ of the SC Nd$_{2-x}$Ce$_x$CuO$_4$ and Pr$_{2-x}$Ce$_x$CuO$_4$ single crystals ($x = 0.13$-0.17) \cite{Takagi1989}. The results are also consistent with a recent core-level spectroscopy study \cite{Horio2017a} where chemical potential of SC Nd$_2$CuO$_4$ thin film was found to be close to that of SC Nd$_{2-x}$Ce$_x$CuO$_4$ ($x=0.15$--0.19) thin films. Fixing the parameters obtained by fitting the Fermi surface ($t'/t = -0.22$, $t''/t' = -0.50$, and $\epsilon_0/t = -0.10$), the constant-energy surface of the SC film at $E = E_\mathrm{F} - 150$ meV could be fitted to the tight-binding model with $t = 0.26$ eV as displayed in Fig.~\ref{PCO_nFS}(a). These values of the parameters $t$, $t'$, $t"$, and $\epsilon_0$ are close to those of SC Nd$_{2-x}$Ce$_x$CuO$_4$ ($x = 0.15$) bulk single crystal ($t = 0.27$ eV, $t'/t = -0.20$, $t''/t' = -0.50$, and $\epsilon_0/t = -0.12$) estimated in a previous ARPES study \cite{Ikeda2009}. Consequently, not only the carrier concentration but also the overall band structure is similar between the Ce-free and Ce-doped T'-type cuprate superconductors.

Since the present Pr$_2$CuO$_4$ thin films do not contain Ce, the variation of the carrier concentration should be attributed to oxygen non-stoichiometry. The shorter $c$-axis length for the SC film indicates that less impurity apical oxygen atoms are incorporated \cite{Radaelli1994,Schultz1996}, but that alone cannot bring the band filling to more than half-filling. The high filling level can be explained only if oxygen vacancies are created in the $Ln_2$O$_2$ layers and/or the CuO$_2$ planes by oxygen reduction and the total amount of oxygen atoms in unit formula becomes less than 4. This assumption is consistent with the experimental observation that $n_\mathrm{FS}$ is larger for the SC films grown under more strongly reducing condition. The electron concentration of $n_\mathrm{FS} = 0.17$ for the SC film can be accounted for if $\sim 2$ \% of total oxygen atoms are deficient. Considering that the $T_\mathrm{c}$ of the Ce-free SC film is even higher than that of optimally Ce-doped thin films \cite{Krockenberger2012,Ikeda2016}, the small amount of oxygen vacancies does not seem to deteriorate the superconductivity.

Recently, using Raman spectroscopy, Kim {\it et al.}~\cite{Kim2017} has observed that apical-oxygen vacancies are inevitable in hole-overdoped La$_{2-x}$Sr$_x$CuO$_4$ thin films, and proposed that their electronic structure has to be reinvestigated taking into account the oxygen vacancies. Oxygen non-stoichiometry is thus a general issue for the cuprates and should not be overlooked. It is important to consider the total amount of doped carriers and oxygen occupancy rather than the amount of chemical substitution alone when discussing the electronic structure of cuprate superconductors.

In conclusion, we have performed ARPES measurements on SC and non-SC Pr$_2$CuO$_4$ thin films and investigated their low-energy electronic structures. The measurements were conducted by protecting the film surfaces with Se cap and removing it before the ARPES measurements. With this method, ARPES measurements on various oxide thin films potentially become possible without the use of an ARPES apparatus connected to a MBE growth chamber. The carrier concentration estimated from the Fermi surface area of the SC Pr$_2$CuO$_4$ thin film was comparable to that of Ce-doped superconductors, and the hopping parameters and the position of the chemical potential which characterize the band structure were also similar between two types of superconductors. The electronic structure of the superconducting cuprates is strongly influenced by oxygen non-stoichiometry, and hence discussion about their physical properties should be based on the actual carrier concentration influenced by oxygen non-stoichiometry rather than the amount of chemical substitution alone.

\begin{acknowledgments}
ARPES experiments were performed under the approval of the Photon Factory Program Advisory Committee (proposal No.~2016G096). This work was supported by Grants-in-Aid from the Japan Society of the Promotion of Science (JSPS) (grant Nos.~14J09200 and 15H02109). M.H. acknowledges financial support from the Advanced Leading Graduate Course for Photon Science (ALPS) and the JSPS Research Fellowship for Young Scientists.
\end{acknowledgments}


\begin{thebibliography}{10}

\bibitem{Tsukada2005a}
A. Tsukada, Y. Krockenberger, M. Noda, H. Yamamoto, D. Manske, L. Alff, and M.
  Naito, Solid State Commun. {\bf 133},  427   (2005).

\bibitem{Matsumoto2009b}
O. Matsumoto, A. Utsuki, A. Tsukada, H. Yamamoto, T. Manabe, and M. Naito,
  Physica C {\bf 469},  924   (2009).

\bibitem{Krockenberger2013}
Y. Krockenberger, H. Irie, O. Matsumoto, K. Yamagami, A. Tsukada, M. Naito, and
  H. Yamamoto, Sci. Rep. {\bf 3},  2235  (2013).

\bibitem{Radaelli1994}
P.~G. Radaelli, J.~D. Jorgensen, A.~J. Schultz, J.~L. Peng, and R.~L. Greene,
  Phys. Rev. B {\bf 49},  15322  (1994).

\bibitem{Schultz1996}
A.~J. Schultz, J.~D. Jorgensen, J.~L. Peng, and R.~L. Greene, Phys. Rev. B {\bf
  53},  5157  (1996).

\bibitem{Xu1996}
X.~Q. Xu, S.~N. Mao, W. Jiang, J.~L. Peng, and R.~L. Greene, Phys. Rev. B {\bf
  53},  871  (1996).

\bibitem{Naito2016}
M. Naito, Y. Krockenberger, A. Ikeda, and H. Yamamoto, Physica C {\bf 523},  28
    (2016).

\bibitem{Das2009}
H. Das and T. Saha-Dasgupta, Phys. Rev. B {\bf 79},  134522  (2009).

\bibitem{Weber2010a}
C. Weber, K. Haule, and G. Kotliar, Nat. Phys. {\bf 6},  574  (2010).

\bibitem{Weber2010b}
C. Weber, K. Haule, and G. Kotliar, Phys. Rev. B {\bf 82},  125107  (2010).

\bibitem{Higgins2006}
J.~S. Higgins, Y. Dagan, M.~C. Barr, B.~D. Weaver, and R.~L. Greene, Phys. Rev.
  B {\bf 73},  104510  (2006).

\bibitem{Mang2004a}
P.~K. Mang, O.~P. Vajk, A. Arvanitaki, J.~W. Lynn, and M. Greven, Phys. Rev.
  Lett. {\bf 93},  027002  (2004).

\bibitem{Adachi2016}
T. Adachi, A. Takahashi, K.~M. Suzuki, M.~A. Baqiya, T. Konno, T. Takamatsu, M.
  Kato, I. Watanabe, A. Koda, M. Miyazaki, R. Kadono, and Y. Koike, J. Phys.
  Soc. Jpn. {\bf 85},  114716  (2016).

\bibitem{Horio2016}
M. Horio, T. Adachi, Y. Mori, A. Takahashi, T. Yoshida, H. Suzuki, L.~C.~C.
  Ambolode, K. Okazaki, K. Ono, H. Kumigashira, H. Anzai, M. Arita, H.
  Namatame, M. Taniguchi, D. Ootsuki, K. Sawada, M. Takahashi, T. Mizokawa, Y.
  Koike, and A. Fujimori, Nat. Commun. {\bf 7},  10567  (2016).

\bibitem{Song2017}
D. Song, G. Han, W. Kyung, J. Seo, S. Cho, B.~S. Kim, M. Arita, K. Shimada, H.
  Namatame, M. Taniguchi, Y. Yoshida, H. Eisaki, S.~R. Park, and C. Kim, Phys.
  Rev. Lett. {\bf 118},  137001  (2017).

\bibitem{Wei2016}
H.~I. Wei, C. Adamo, E.~A. Nowadnick, E.~B. Lochocki, S. Chatterjee, J.~P. Ruf,
  M.~R. Beasley, D.~G. Schlom, and K.~M. Shen, Phys. Rev. Lett. {\bf 117},
  147002  (2016).

\bibitem{Riou2004}
G. Riou, P. Richard, S. Jandl, M. Poirier, P. Fournier, V. Nekvasil, S.~N.
  Barilo, and L.~A. Kurnevich, Phys. Rev. B {\bf 69},  024511  (2004).

\bibitem{Richard2004}
P. Richard, G. Riou, I. Hetel, S. Jandl, M. Poirier, and P. Fournier, Phys.
  Rev. B {\bf 70},  064513  (2004).
  
\bibitem{Yu2007}
W. Yu, B. Liang, P. Li, S. Fujino, T. Murakami, I. Takeuchi, and
R. L. Greene, Phys. Rev. B {\bf 75},  020503(R)  (2007).

\bibitem{Horio2017a}
M. Horio, Y. Krockenberger, K. Yamamoto, Y. Yokoyama, K. Takubo, Y. Hirata, S.
  Sakamoto, K. Koshiishi, A. Yasui, E. Ikenaga, S. Shin, H. Yamamoto, H.
  Wadati, and A. Fujimori, Phys. Rev. Lett. {\bf 120},  257001  (2018).

\bibitem{Richard2007}
P. Richard, M. Neupane, Y.-M. Xu, P. Fournier, S. Li, P. Dai, Z. Wang, and H.
  Ding, Phys. Rev. Lett. {\bf 99},  157002  (2007).

\bibitem{Song2012}
D. Song, S.~R. Park, C. Kim, Y. Kim, C. Leem, S. Choi, W. Jung, Y. Koh, G. Han,
  Y. Yoshida, H. Eisaki, D.~H. Lu, Z.-X. Shen, and C. Kim, Phys. Rev. B {\bf
  86},  144520  (2012).

\bibitem{Ikeda2009}
M. Ikeda, T. Yoshida, A. Fujimori, M. Kubota, K. Ono, H. Das, T. Saha-Dasgupta,
  K. Unozawa, Y. Kaga, T. Sasagawa, and H. Takagi, Phys. Rev. B {\bf 80},
  014510  (2009).

\bibitem{Sakakibara2010}
H. Sakakibara, H. Usui, K. Kuroki, R. Arita, and H. Aoki, Phys. Rev. Lett. {\bf
  105},  057003  (2010).

\bibitem{Krockenberger2015}
Y. Krockenberger, M. Horio, H. Irie, A. Fujimori, and H. Yamamoto, Applied
  Physics Express {\bf 8},  053101  (2015).

\bibitem{Xu2012}
S.-Y. Xu, M. Neupane, C. Liu, D. Zhang, A. Richardella, L. Andrew~Wray, N.
  Alidoust, M. Leandersson, T. Balasubramanian, J. Sanchez-Barriga, O. Rader,
  G. Landolt, B. Slomski, J. Hugo~Dil, J. Osterwalder, T.-R. Chang, H.-T. Jeng,
  H. Lin, A. Bansil, N. Samarth, and M. Zahid~Hasan, Nat. Phys. {\bf 8},  616
  (2012).

\bibitem{Liu2012}
D. Liu, W. Zhang, D. Mou, J. He, Y.-B. Ou, Q.-Y. Wang, Z. Li, L. Wang, L. Zhao,
  S. He, Y. Peng, X. Liu, C. Chen, L. Yu, G. Liu, X. Dong, J. Zhang, C. Chen,
  Z. Xu, J. Hu, X. Chen, X. Ma, Q. Xue, and X. Zhou, Nat. Commun. {\bf 3},  931
   (2012).

\bibitem{Suzuki1990a}
T. Suzuki, M. Nagoshi, Y. Fukuda, K. Oh-ishi, Y. Syono, and M. Tachiki, Phys.
  Rev. B {\bf 42},  4263  (1990).

\bibitem{Armitage2001b}
N.~P. Armitage, D.~H. Lu, C. Kim, A. Damascelli, K.~M. Shen, F. Ronning, D.~L.
  Feng, P. Bogdanov, Z.-X. Shen, Y. Onose, Y. Taguchi, Y. Tokura, P.~K. Mang,
  N. Kaneko, and M. Greven, Phys. Rev. Lett. {\bf 87},  147003  (2001).

\bibitem{Matsui2005a}
H. Matsui, K. Terashima, T. Sato, T. Takahashi, S.-C. Wang, H.-B. Yang, H.
  Ding, T. Uefuji, and K. Yamada, Phys. Rev. Lett. {\bf 94},  047005  (2005).

\bibitem{Park2007}
S.~R. Park, Y.~S. Roh, Y.~K. Yoon, C.~S. Leem, J.~H. Kim, B.~J. Kim, H. Koh, H.
  Eisaki, N.~P. Armitage, and C. Kim, Phys. Rev. B {\bf 75},  060501  (2007).

\bibitem{Park2013}
S.~R. Park, T. Morinari, D.~J. Song, C.~S. Leem, C. Kim, S.~K. Choi, K. Choi,
  J.~H. Kim, F. Schmitt, S.~K. Mo, D.~H. Lu, Z.-X. Shen, H. Eisaki, T. Tohyama,
  J.~H. Han, and C. Kim, Phys. Rev. B {\bf 87},  174527  (2013).
  
\bibitem{Suppl}
See Supplemental Material for more detailed analyses on the AFM pseudogap and the constant-energy surfaces.

\bibitem{Kaminski2004}
A. Kaminski, S. Rosenkranz, H.~M. Fretwell, J. Mesot, M. Randeria, J.~C.
  Campuzano, M.~R. Norman, Z.~Z. Li, H. Raffy, T. Sato, T. Takahashi, and K.
  Kadowaki, Phys. Rev. B {\bf 69},  212509  (2004).

\bibitem{Matt2015}
C.~E. Matt, C.~G. Fatuzzo, Y. Sassa, M. M\aa{}nsson, S. Fatale, V. Bitetta, X.
  Shi, S. Pailh\`es, M.~H. Berntsen, T. Kurosawa, M. Oda, N. Momono, O.~J.
  Lipscombe, S.~M. Hayden, J.-Q. Yan, J.-S. Zhou, J.~B. Goodenough, S. Pyon, T.
  Takayama, H. Takagi, L. Patthey, A. Bendounan, E. Razzoli, M. Shi, N.~C.
  Plumb, M. Radovic, M. Grioni, J. Mesot, O. Tjernberg, and J. Chang, Phys.
  Rev. B {\bf 92},  134524  (2015).

\bibitem{Takagi1989}
H. Takagi, S. Uchida, and Y. Tokura, Phys. Rev. Lett. {\bf 62},  1197  (1989).

\bibitem{Krockenberger2012}
Y. Krockenberger, H. Yamamoto, A. Tsukada, M. Mitsuhashi, and M. Naito, Phys.
  Rev. B {\bf 85},  184502  (2012).

\bibitem{Ikeda2016}
A. Ikeda, H. Irie, H. Yamamoto, and Y. Krockenberger, Phys. Rev. B {\bf 94},
  054513  (2016).

\bibitem{Kim2017}
G. Kim, G. Christiani, G. Logvenov, S. Choi, H.-H. Kim, M. Minola, and B.
  Keimer, Phys. Rev. Materials {\bf 1},  054801  (2017).

\end{thebibliography}

\end{document}


\title{Supplementary Information \\ \vspace{1cm} ARPES studies on the low-energy electronic structure of superconducting Pr$_2$CuO$_4$ driven by oxygen non-stoichiometry \vspace{1cm}}

\author{M. Horio$^1$, Y. Krockenberger$^2$, K. Koshiishi$^1$, S. Nakata$^1$, K. Hagiwara$^1$, M. Kobayashi$^3$, K. Horiba$^3$, H. Kumigashira$^3$, H. Irie$^2$, H. Yamamoto$^2$, and A. Fujimori$^1$}

\begingroup
\let\clearpage\relax
\let\vfil\relax
\maketitle
\endgroup

\noindent
{\it \footnotesize $^1$Department of Physics, University of Tokyo, Bunkyo-ku, Tokyo 113-0033, Japan} \\
{\it \footnotesize $^2$NTT Basic Research Laboratories, NTT Corporation, Atsugi, Kanagawa 243-0198, Japan} \\
{\it \footnotesize $^3$KEK, Photon Factory, Tsukuba 305-0801, Japan} \\
\vfill

\noindent
$^*$e-mail: horio@wyvern.phys.s.u-tokyo.ac.jp \\

\newpage

\section{Suppression of the antiferromagnetic pseudogap from energy distribution curves}
Energy distribution curves (EDCs) of the superconducting (SC) Pr$_2$CuO$_4$ film at several $k_\mathrm{F}$ positions are plotted from the nodal to antinodal regions and in Supplementary Fig.~\ref{EDC}(a). Each spectrum has been normalized to the acquisition time and the photon flux. The spectral intensity is apparently rather weak around the node, and the peak intensity develops on approaching the antinode. The EDCs contain not only the intrinsic photoemission signals but also momentum-independent background produced by scattered secondary photoelectrons. First, we constructed the background EDC by integrating the EDCs in the gray shaded region in the inset of Supplementary Fig.~\ref{EDC}(a) and normalizing to the acquisition time and the photon flux, as empirically employed for other cuprates such as Bi$_2$Sr$_2$CaCu$_2$O$_{8+x}$ \cite{Kaminski2004} and La$_{2-x}$Sr$_x$CuO$_4$ \cite{Matt2015}, and subtracted it from the EDCs at the $k_\mathrm{F}$ positions. Assuming that photoionization matrix element for the background does not vary strongly with momentum \cite{Kaminski2004}, the spectral intensity of the background at a given binding energy was kept constant in the entire momentum regions considered here. After the EDCs have been normalized to the intensity at high binding energies ($-0.4$ eV $\leq  E - E_\mathrm{F} \leq -0.25$ eV) along with the background subtraction, a sharp quasi-particle peak becomes visible on the entire Fermi surface as shown in Supplementary Fig.~\ref{EDC}(b). Although around the hot spot a two-peak structure indicative of antiferromagnetic (AFM) band splitting [Supplementary Fig.~\ref{EDC}(b)] and a slight suppression of the peak intensity within $E - E_\mathrm{F} = \pm 20$ meV [Supplementary Figs.~\ref{EDC}(c),(d)] are observed, there is no clear gap anywhere on the Fermi surface, suggesting a strong suppression of the AFM pseudogap.

\begin{figure}[h]
\begin{center}
\includegraphics[width=0.9\textwidth]{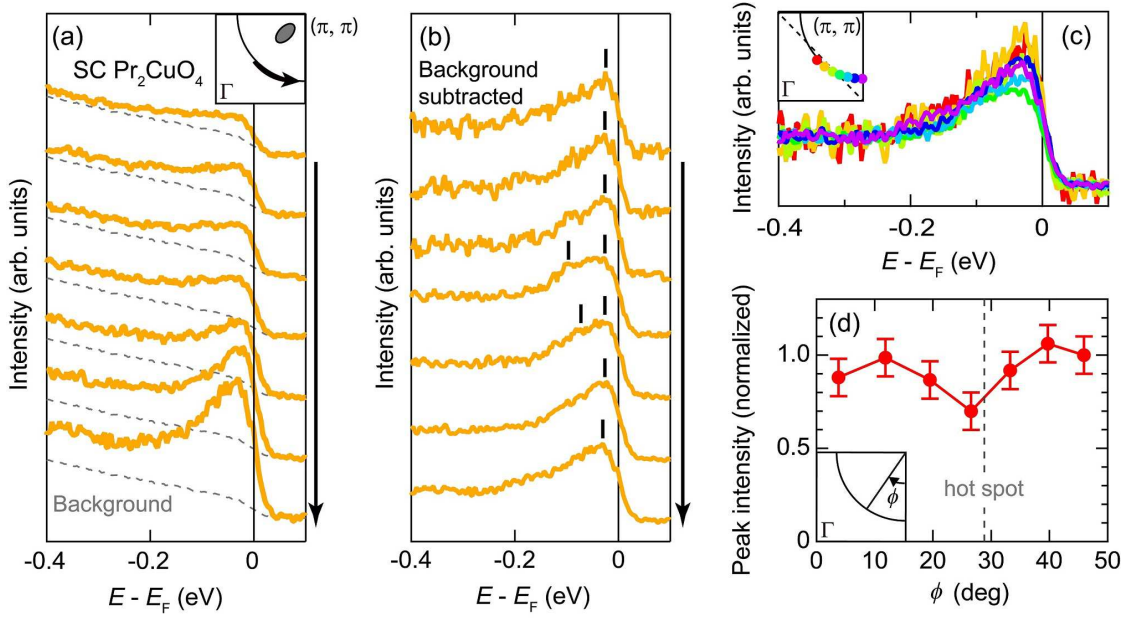}
\end{center}
\caption{Energy distribution curves (EDCs) on the Fermi surface of the SC Pr$_2$CuO$_4$ sample. (a) EDCs of the SC Pr$_2$CuO$_4$ thin film at several $k_\mathrm{F}$ positions plotted from the nodal to antinodal regions. Background has been taken from the gray shaded region in the inset and plotted with dashed curves. (b) Background-subtracted EDCs. The intensity has been normalized to that integrated within $-0.4$ eV $\leq  E - E_\mathrm{F} \leq -0.25$ eV after background subtraction. Peak positions are indicated by vertical bars. (c) The same EDCs as those in (b) but plotted without offsets. The inset shows the corresponding momentum positions. (d) Momentum dependence of the peak intensity integrated within $E - E_\mathrm{F} = \pm 20$ meV. The values have been normalized to that at the nodal point. The Fermi surface angle $\phi$ is defined in the inset.}
\label{EDC}
\end{figure}

\section{Determination of constant-energy surfaces at $E = E_\mathrm{F} - 150$ meV}
\begin{figure}[h]
\begin{center}
\includegraphics[width=0.9\textwidth]{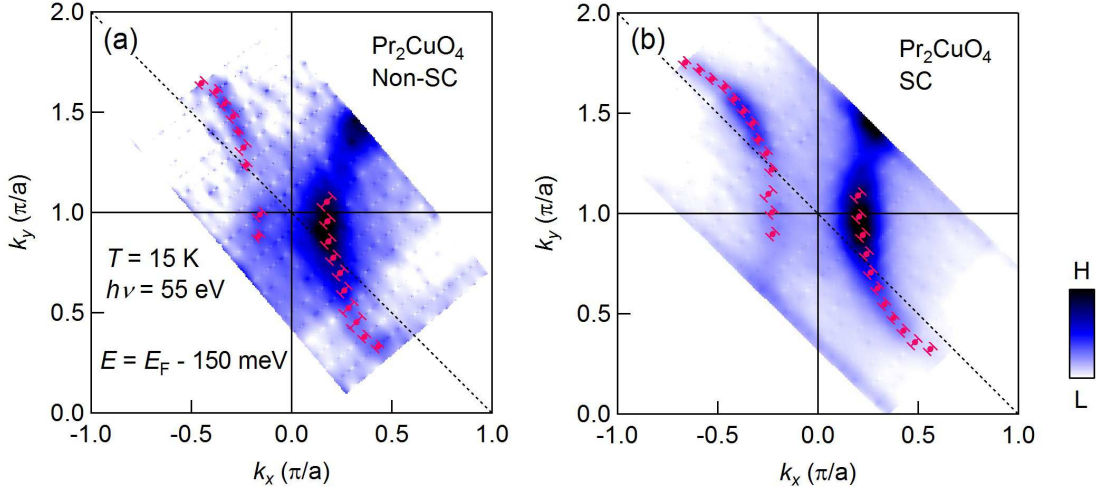}
\end{center}
\caption{Constant-energy surfaces of the Pr$_2$CuO$_4$ thin films at $E = E_\mathrm{F} - 150$ meV. (a) Non-SC Pr$_2$CuO$_4$. (b) SC Pr$_2$CuO$_4$. Pink dots represent the peak positions of MDCs integrated within $\pm 10$ meV with respect to $E = E_\mathrm{F} - 150$ meV.}
\label{Const150}
\end{figure}

Supplementary Fig.~\ref{Const150} shows a constant-energy surface mapping at $E = E_\mathrm{F} - 150$ meV. Each momentum distribution curve (MDC) integrated within $E = E_\mathrm{F} - 150 \pm 10$ meV is fitted to a Lorentzian, and thus determined peak positions are plotted as pink dots. The error bar corresponds to a quarter of the half width at half maximum (HWHM) of the Lorentzian. The MDC peaks obtained in the wide momentum region ranging from the first to second Brillouin zones shown in Supplementary Fig.~\ref{Const150} are folded to a quarter of the Brillouin zone by symmetry operation, and then symmetrized with respect to the $\Gamma$-($\pi$, $\pi$) line to reproduce the constant-energy surfaces shown in Fig.~4(a) of the main text.

\section{Determination of Fermi surfaces}
In Supplementary Fig.~\ref{FS}, Fermi surfaces are mapped and estimated $k_\mathrm{F}$ positions are plotted. In the entire momentum region for the SC film and in the antinodal region for the non-SC film, where the spectral intensity near $E_\mathrm{F}$ is high enough, $k_\mathrm{F}$ positions are determined from the peak positions of momentum distribution curves (MDCs) integrated within $E_\mathrm{F} \pm 10$ meV, and plotted as pink dots in Supplementary Figs.~\ref{FS}(a) and (b). The error bar represents a quarter of HWHM of the Lorentzian. Since the spectral intensity near $E_\mathrm{F}$ is low around the node and hot spot for the non-SC film, and the MDCs at $E_\mathrm{F}$ does not exhibit a clear peak, $k_\mathrm{F}$ positions in that momentum region are determined by extrapolating MDC peaks from $E - E_\mathrm{F} = -70$ meV to $-20$ meV towards $E_\mathrm{F}$ assuming a linear band dispersion. Thus determined $k_\mathrm{F}$ points are plotted as green dots in Supplementary Fig.~\ref{FS}(a). The magnitude of the experimental error was defined as $2\sigma$ of the extrapolation. Fermi surfaces plotted in Fig.~4(b) of the main text consist of those $k_\mathrm{F}$ points symmetrized using the same method as described above.

\begin{figure}[h]
\begin{center}
\includegraphics[width=0.9\textwidth]{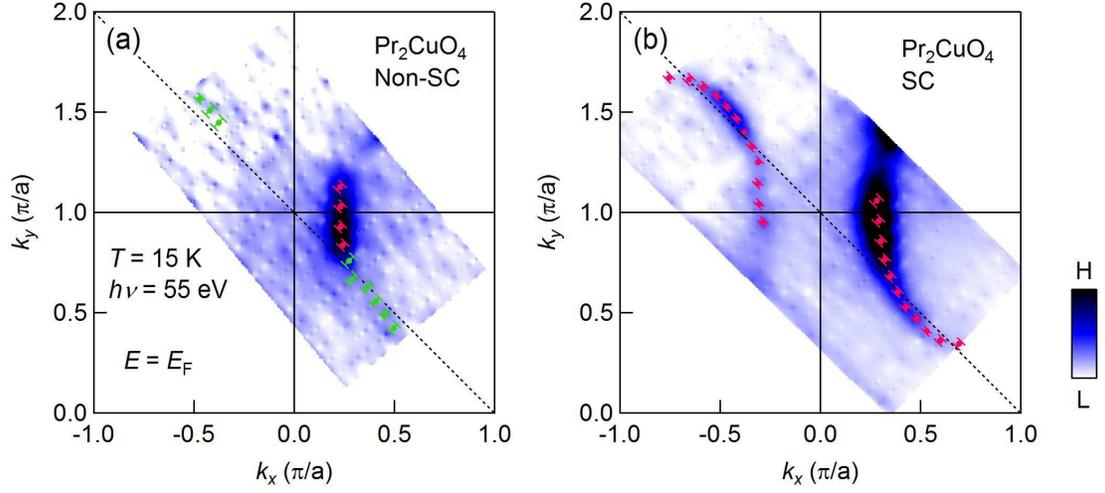}
\end{center}
\caption{Fermi surfaces of Pr$_2$CuO$_4$ thin films. (a) Non-SC Pr$_2$CuO$_4$. (b) SC Pr$_2$CuO$_4$. Pink dots represent $k_\mathrm{F}$ points which are determined from the peak positions of MDCs integrated within $E_\mathrm{F} \pm 10$ meV. In the nodal region of the non-SC film, $k_\mathrm{F}$ position are determined by fitting MDC peaks within $E - E_\mathrm{F} = -70$ meV to $-20$ meV to a line and extrapolating it to $E_\mathrm{F}$, and are plotted as green dots.}
\label{FS}
\end{figure}
